\documentclass[conference]{IEEEtran}
\ifCLASSINFOpdf
\else
\fi
\newcommand{\quo}{\frac{d_l}{d_r}}
\newcommand{\field}[1]{\mathbb{#1}}

\usepackage[T1]{fontenc}
\usepackage{amsmath,amssymb,amsthm}

\newtheorem{thm}{Theorem}[section]
\newtheorem{prop}{Proposition}[section]

\newtheorem{lem}[prop]{Lemma}

\newtheorem*{claim}{Claim}

\newtheorem{cor}[prop]{Corollary}

\theoremstyle{definition}
\newtheorem{defn}[prop]{Definition}
\newtheorem{ex}{Example}

\newtheorem*{rem}{Remark}
\newtheorem*{notation}{Notation}
\newtheorem*{disc}{Discussion}

\begin{document}
%
\title{Extended Extremes of Information Combining}

\author{\IEEEauthorblockN{Lucas Boczkowski}
\IEEEauthorblockA{Ecole Normale Superieure\\
Paris, France\\ 75005\\
Email: lucas.boczkowski@ens.fr}
}

\maketitle

\begin{abstract}
Extremes of information combining inequalities play an important
role in the analysis of sparse-graph codes under message-passing
decoding.  We introduce new tools for the derivation of such
inequalities, and show by means of a concrete examples how they can be
applied to solve some optimization problems in the analysis of
low-density parity-check codes.  \end{abstract}



%
\IEEEpeerreviewmaketitle

\section{Setting}
In order to understand iterative decoding of low-density-parity-check
codes (LDPC), two operations need to be studied. These operations
are the variable node convolution $\otimes$ and the check node
convolution $\boxtimes$. They correspond to the merging of information
respectively by variable nodes and by check nodes in the iterative
decoding process. The reader is assumed to be familiar with LDPC
codes as well as the formalism of modeling channels by densities.
A very complete introduction to the topic is \cite{mct}.

The notion of extremes of information combining (EIC) was introduced
by I. Land, P. Hoeher, S. Huettinger, and J. B. Huber in \cite{EIC03}, 
and further extended by I. Sutskover, S. Shamai, and J. Ziv, see \cite{EIC05} or \cite{EIC07}.  The
idea of EIC is to associate to densities certain functionals, e.g.
the entropy functional, and to see how these functionals behave
under the combining of information, i.e. the two kinds of convolutions.
The purpose of this work is to solve optimizing problems that arise
in this setting. We will focus solely on the check convolution
$\boxtimes$ although many statements can be proven in the same way
for the variable node convolution.
\subsection{Notations}
There are several representations for a binary memoryless and symmetric-output channel (BMS). 
As is done for instance in \cite{mct}, we see a BMS as a convex combination of binary symmetric channels (BSC), 
given by a weight distribution $w$. Then we have (by definition)
\begin{ex}[Binary Symmetric Channel BSC($\epsilon$)]
\begin{align*}
w_{\text{BSC($\epsilon$)}} = \delta_{\epsilon}
\end{align*}
\end{ex}

\begin{ex}[Binary Erasure Channel BEC($\epsilon$)]
\begin{align*}
w_{\text{BEC($\epsilon$)}} = \bar{\epsilon}\delta_{0} + \epsilon\delta_{\frac{1}{2}}
\end{align*}
\end{ex}

The functionals of interest in this domain are 
\begin{align*}
E(a) &= \int_0^{\frac{1}{2}} dw_a(\epsilon) \epsilon \\
H(a) &= \int_0^{\frac{1}{2}} dw_a(\epsilon) h_2\big(\epsilon\big) \\
B(a) &= \int_0^{\frac{1}{2}} dw_a(\epsilon) 2\sqrt{\epsilon (1-\epsilon)} 
\end{align*}
which we call respectively the error probability, the entropy and
the Battacharyya functional. These can all be thought of as measures
of the channel quality. They are equal to $0$ for the perfect
channel and equal to $1$ for a useless channel. Applying these functionals
to the check convolution of two densities corresponds to
\begin{align*}
E(a \boxtimes b) &= \frac{1-(1-2E(a))(1-2E(b))}{2} \\
H(a\boxtimes b) &= \int dw_a(\epsilon) dw_b(\epsilon') h_2\Big(\frac{1-(1-2\epsilon)(1-2\epsilon')}{2}\Big)\\
B(a\boxtimes b) &= \int dw_a(\epsilon) dw_b(\epsilon')\sqrt{1-\big( (1-2\epsilon)(1-2\epsilon') \big)^2}
\end{align*}
In the sequel we will frequently refer to the following two functions, $f_H$ and $f_B$ :
\begin{align*}
f_H : X &\in [0;1] \mapsto h_2\big(\frac{1-X}{2}\big) \\
f_B : X &\in [0;1] \mapsto \sqrt{1-X^2}
\end{align*}

\subsection{Motivation}
A classical result in EIC, shown in \cite{EIC05} and \cite{EIC03},
is the following.
\begin{thm}\label{prev}
Let $b_0$ be any BMS channel. Amongst channels $a$, with fixed entropy $H$, $H(a \boxtimes b_0)$ is
\begin{itemize}
\item minimized by the \text{BSC}$(h_2^{-1}(h))$ and
\item maximized by the \text{BEC}(h).
\end{itemize}
\end{thm}
A quick and useful application of \ref{prev} is to give bounds on
the thresholds of LDPC codes. The same statement can be done with
the Battacharyya functional $B$. We will derive an alternate (calculus
free) proof of the second item in \ref{obounds}.

Sometimes one might need to deal with non-linear
expressions such as $H(a^{\boxtimes 4})-H(a^{\boxtimes 12})$. Let
us sketch very loosely, following  \cite{URK11.2},  how such
expressions can appear. Apart from the Shannon threshold another
threshold called the \emph{Area Threshold} can be defined. The Area
threshold depends on the code and channel family under consideration.
In the case of a code taken from the $(d_l,d_r)$ regular ensemble,
one can compute this threshold $h^A$.

Consider a code taken from the $(d_l,d_r)$ regular ensemble, and transmission over a "gentle" channel family $\{ c_{\sigma}\}_{\underline{\sigma}}^{\overline{\sigma}}$, that is a family that is smooth, ordered, and complete\footnote{The definitions of these terms can be found for instance in \cite{mct}. Examples of such families include amongst many others the $\{\text{BEC}(h)\}_0^{1}$ and the $\{\text{BSC}(h)_0^{1}$ as well as combinations of these two, and other classical families like the $\{\text{BAWGNC}(\sigma)\}_0^{\infty}$.}. Ordered means that the bigger the channel parameter $\sigma$ the worst the channel is, in other words, all the functionals introduced above increase with $\sigma$, "smooth" means we can derivate $\sigma$ in the integrals.

Then one can define a GEXIT curve in the following manner. Take a FP $(c_{\sigma},x)$ and define $y = x^{\boxtimes d_r-1}$. Then plot,
$$
\big(H(c_{\sigma}), G(c_{\sigma},y^{\otimes d_l}) \big).
$$
Here
$
G(c_{\sigma}, \cdot) = \frac{H(\frac{dc_{\sigma}}{d\sigma} \otimes \cdot)}{H(\frac{c_{\sigma}}{d\sigma})}
$
In the case of the BEC, changing the channel parameter $\sigma$, corresponds to revealing certain bits, and the kernel $G(c_{\sigma},\cdot)$ represents the probability that this bit was not previously known from the observation of the value of other neighboring bits\footnote{Neighbors is to be understood in the sense of the Tanner graph as usual.}. In general everything has the same meaning but with soft information. 

The kernel models how much more (compared to if we use only extrinsic observations) information is known about a generic bit, if the channel is made slightly better. For instance if the channel changes from being useless ($H(c)=1$) to slightly better, all the information we get is useful because with a useless channel nothing is known. So there is a point at $(1,1)$. 

So intuitively, the area below this curve (assuming it exists and is smooth) between $h$ and $1$, should be a measure in bits of the total useful information that we get through BP decoding for $\sigma$ s.t. $H(c_{\sigma})=h$. As the rate of the code is roughly $1-\frac{d_l}{d_r}$, $1-\frac{d_l}{d_r}$ bits of information is enough to fully determine a codeword.

It is then natural to define the Area threshold $h^{A}$ as the point on the horizontal axis s.t. the area below the curve starting at $h^{A}$ to $1$ is equal to the design rate $1-\frac{d_l}{d_r}$. Of course this notion is dependent on the channel family.

However, an iterative decoder like BP might not be able to "use" all this information\footnote{Think of the BEC for which what BP does, is solving a system of equations by iteratively solving equations where all variables are known but one. Even if the system is full rank there might still be large portions that remain unknown to BP.}. So in general $h^{\text{BP}} \leq h^{A}$.

It would make sense that $h^{\text{MAP}} = h^{A}$, although in the general setting all that is known is $h^{\text{MAP}} \leq h^{A}$

In \cite{URK11.2}, it is shown that the value of the integral from $h$ to $1$ is
\begin{align}
1-\frac{d_l}{d_r} - h-(d_l-1-\frac{d_l}{d_r})H(x^{\boxtimes d_r}) + (d_l-1)H(x^{\boxtimes d_r-1})
\end{align}
where $x$ is "the" BP fixed point with entropy $h$ for the channel family under consideration.


The value of $h^{A}$, turns out to be the right bound of the domain where the following holds
\begin{align}\label{area}
- h-(d_l-1-\frac{d_l}{d_r})H(x^{\boxtimes d_r}) + (d_l-1)H(x^{\boxtimes d_r-1}) \geq 0.
\end{align}
Here, $x$ is "the" density evolution fixed point with entropy $h$, using
belief propagation (BP) decoding. In \cite{URK11.2} it is shown
that indeed (\ref{area}) holds, \emph{universally} over all BMS
channels $x$ with entropy lower or equal to $\frac{d_l}{d_r}$, in
the asymptotic regime $d_l,d_r \to \infty$ with $\quo$ fixed. This
implies the Area threshold universally approaches the Shannon
threshold. We will derive another proof of this fact in Section
\ref{sareat} (see Proposition \ref{cclarea}).

In \cite{URK11.2} it is then shown that a class of spatially coupled
codes achieve the Area threshold, under BP decoding. This combined
with the fact above, gives a new way to achieve capacity.

\section{Results}
Our results fit in a slightly more general framework than that of Theorem \ref{prev}: 
we will consider expressions of the type $\Phi(\rho(a))$ 
where $\rho$ is a polynomial, and $\Phi$ is either $H$ or $B$. 
We use the following notation
\begin{notation}
Let $\rho(X) = \sum c_i X^{i}$ be any polynomial s.t. $\rho(0) =0$,\footnote{Instead of considering polynomials who vanish at $0$, we could use a convention like $a^{\boxtimes 0}=$"Perfect Channel".} i.e. $c_0 = 0$ and $\Phi$ be one of the functionals above. We will use the convention
\begin{align}\label{def1}
\Phi(\rho(a)) \stackrel{\text{def}}{=} \sum_i c_i \Phi(a^{\boxtimes i})
\end{align}
\end{notation}

The following two statements are our main results. 
We prove them in the next section.
\begin{prop}\label{t1}
Let $\rho$ be any polynomial s.t. $\rho(0)=0$, 
and $\Phi$ be $H$ or $B$. Consider the following problem
\begin{align*}
\mbox{\text{MAX} }&\Phi(\rho(a))\\
\mbox{s.t. }&\Phi(a) = \phi_0
\end{align*}
Then, if $\rho$ is $\cup$-convex over $[0;f_{\Phi}^{-1}(\phi_0)^2]$, the BEC solves this problem.
\end{prop}

\begin{prop}\label{t2}
Let $\rho$ be any polynomial s.t. $\rho(0)=0$, and $\Phi$ be $H$ or $B$. Consider the following problem
\begin{align*}
\mbox{\text{OPT} }&\Phi(\rho(a))\\
\mbox{s.t. }&E(a) = \epsilon
\end{align*}
Then, if $\rho$ is increasing over $[0;1-2\epsilon]$,
\begin{itemize}
\item the \text{BEC} minimizes this problem.
\item the \text{BSC} maximizes this problem.
\end{itemize}
\end{prop}
\begin{disc}
The hypotheses for these propositions are probably not tight, they just ease the proofs. 
The reader should not pay too much attention to the obscure terms $f_{\Phi}^{-1}(\phi_0)^2$.

The maximizing part in the previous result \ref{prev} follow as a special case of Proposition \ref{t1} with $\rho=X^d$.
Our improvement, technically speaking, is dealing with other polynomials than $X^d$.
\end{disc}
Proposition \ref{t1} only addresses half of the question. We suspect that 
in most cases the minimizer is the BSC, and pose this as an interesting open question. 
Dealing with the problem \ref{area} requires that we have a lower bound. 
This is the purpouse of the following lemma

\begin{lem}\label{l1}
Suppose $\rho$ is increasing over $[0;f_{\Phi}^{-1}(\phi_0)^2]$. Then, for all channels $a$ with $\Phi(a)=\phi_0$,
\begin{align}\label{b1}
\Phi(\rho(a)) \geq \rho(1) - \rho(f_{\Phi}^{-1}(\phi_0)^2)
\end{align}
\end{lem}
\vspace*{0.1cm}
\section{Proofs}
Before we start the proof, a few preliminary observations are needed.
\subsection{Preliminary observations}\label{prelim}
Let $\Phi$ be either $H$, the entropy or $B$, the Battacharyya functional. In both cases the "kernel" $f_{\Phi}$ can be expanded in power series,
\begin{align*}
f_\Phi(X) = 1 - \sum_{1}^{\infty} a_{\Phi,n} X^{2n}
\end{align*}
where equality still holds for $X=1$. The crucial property of
$\big(a_{\Phi,n}\big)_n$ is that all the terms are positive and
furthermore
$$
\sum_{n\geq0} a_{\Phi,n} = 1
$$
The explicit formulas are 
\begin{align*}
a_{H,n} &= \frac{1}{2\log(2)n(2n-1)} \\
a_{B,n} &= \frac{{2n\choose n}}{(2n-1)4^n}
\end{align*}
This expansion can be plugged in the definition of $\Phi(a)$ to yield
\begin{align*}
1-\Phi(a) &= 1-\int dw_a(\epsilon)f_{\Phi}(1-2\epsilon)\\
&=\sum a_{\Phi,n} \int dw_a(\epsilon)(1-2\epsilon)^{2n}
\end{align*}
and we can proceed in a similar fashion for $\Phi(a \boxtimes b)$ or more complicated expressions.
\begin{defn}[moments]
For a channel $a$, its $n$-th moment is defined by
$$\gamma_{a,n} = \int dw_a(\epsilon)(1-2\epsilon)^{2n}.$$ 
\end{defn}
We call the $\gamma_{a,n}$s \emph{moments} even if, strictly speaking,
they are not. Note that in terms of moments, the \text{BEC} is
characterized by having all its moments equal, and the \text{BSC}
by having moments that decrease geometrically.
\begin{ex}
Fix $\Phi = \phi_0$, where $\Phi$ is either the Battacharyya functional or the entropy. 
Consider the \text{BEC} and the \text{BSC} s.t. there $\Phi(.)$ is equal to $\phi_0$. Then,
\begin{align*}
\gamma_{\text{BEC},n} &= 1-\phi_0 \\
\gamma_{\text{BSC},n} &= f_{\Phi}^{-1}(\phi_0)^{2n}
\end{align*}
\end{ex}
With this definition
\begin{align}\label{w1}
1-\Phi(a) = \sum_{n\geq 1} a_{\Phi,n} \gamma_{a,n}
\end{align}

Note also that if $\Phi=H$, then $1-\Phi$ is no other than $C$, the capacity functional. Also, using Fubini, we see that 
$
 \int dw_a(\epsilon) dw_b(\epsilon')(1-2\epsilon)^{2n}(1-2\epsilon')^{2n} = \gamma_{a,n} \gamma_{b,n}
$
and it follows that
\begin{align}{\label{e1}}
1-\Phi(a\boxtimes b) = \sum a_{\Phi,n} \gamma_{a,n} \gamma_{b,n}
\end{align}
and this yields straightforwardly
\begin{align}{\label{f1}}
1-\Phi(a^{\boxtimes i}) = \sum a_{\Phi,n} \gamma_{a,n}^{i}
\end{align}
More generally, if $\rho=\sum_{i \geq 1} c_i X^{i}$ is a polynomial
\begin{align*}
\Phi(\rho(a)) &\stackrel{(\ref{def1})}{=} \sum_i c_i \Phi(a^{\boxtimes i}) \\
&\stackrel{(\ref{f1})}{=}\sum_i c_i \left( 1-\sum_n (a_{\Phi,n} \gamma_{a,n}^i)\right) \\
&\stackrel{\text{def}}{=} \sum_i c_i-\sum_n a_{\Phi,n} \underbrace{\sum_i c_i \gamma_{a,n}^i}_{\rho(\gamma_{a,n})}
\end{align*}
which can be rewritten as
\begin{align}\label{phirhogen}
\Phi(\rho(a)) = \rho(1) - \sum_n a_{\Phi,n}\rho(\gamma_{a,n})
\end{align}
Although very simple, the expansion above gives an efficient way
to derive numerous bounds. All the proofs presented here rely heavily
on it.

It will be convenient in the sequel to know the range
the moments can achieve. They are decreasing and positive. So the
biggest moment is the first $\gamma_{a,1}$. The next lemma states
what channel $a$ maximizes $\gamma_{a,1}$.

\begin{lem}\label{l2}
Amongst all channels $a$, s.t. $\Phi(a) = \phi_0$, the \text{BSC} maximizes $\gamma_{a,1}$.
\end{lem}
\begin{proof}
The function $x \mapsto x^n$ is $\cup$-convex. Using Jensen's inequality  
\begin{align*}
\gamma_{a,n} = \int dw_a(\epsilon) (1-2\epsilon)^{2n} \geq \Big(\int dw_a(\epsilon) (1-2\epsilon)^2\Big)^n = \gamma_{a,1}^n
\end{align*}
Then notice
\begin{align*}
1-\phi_{0} = \sum a_{\Phi,n} \gamma_{a,n} &\geq \sum a_{\Phi,n} \gamma_{a,1}^n \\
&=1-f_{\Phi}(\sqrt{\gamma_{a,1}})
\end{align*}
Inverting this inequality - $f_{\Phi}^{-1}$ is decreasing because $f_{\Phi}$ is - gives
\begin{align*}
\gamma_{a,1} \leq (f_{\Phi}^{-1}(\phi_0))^2
\end{align*}
The bound is attained by and only by the \text{BSC}, for which indeed
\begin{align}\label{g1bsc}
\gamma_{\text{BSC},1} = f_{\Phi}^{-1}(\phi_0)^2
\end{align}
\end{proof}

\begin{notation}
We may write $\gamma_1$ instead of $\gamma_{\text{BSC},1}$.
\end{notation}
Bounds can be used at two different levels. Either we bound the moments themselves - like in the derivation of \ref{l1} - that would be the first level. Or we can look at the expressions from one step further and see
$
\sum a_{\Phi,n} \gamma_{a,n}
$
as an expectation 
$
E(\gamma)
$.
Here the expectation is taken w.r.t to a discrete measure given
by the weights $(a_{\Phi,n})$. In this second setup, we can then use classical inequalities, like 
the Jensen inequality. That is the idea of the proof of \ref{t1}

\subsection{Proof of \ref{t1}}\label{pt1}
Notice, by assumption and Lemma \ref{l1} the range over which $\rho$ is convex covers the values the moments can take.
\begin{align*}
\Phi(\rho(a)) &\stackrel{(\ref{phirhogen})}{=} \rho(1) - \sum_n a_{\Phi,n}\rho(\gamma_{a,n}) \\
&\stackrel{\text{Jensen}}{\leq} \rho(1) - \rho\big( \sum_n a_{\Phi,n}\gamma_{a,n}\big)\\
&\stackrel{(\ref{w1})}{=}\rho(1) - \rho(1-\phi_0)
\end{align*}
To conclude notice that 
$
\rho(1) - \rho(1-\phi_0) = \Phi(\rho(\text{BEC}(\phi_0)))
$.

\subsection{Proof of \ref{t2}}
Proposition \ref{t2} is a direct corollary of 
\begin{lem}\label{l2}
For all $n\in\field{N}$, amongst the channels $a$ with fixed $E$ say $\epsilon$, the one who minimizes (resp. maximizes) $\gamma_{a,n}$ is the \text{BSC}$(\epsilon)$ (resp \text{BEC}$(2\epsilon)$).
\end{lem}
\begin{proof}[Proof of \ref{l2}]
Even though it is not mandatory to do so, we can choose according to Caratheodory Prinicple (see \ref{conv}) to restrict ourselves to combinations of two $\delta$'s.
$$ a = \alpha \delta_{\epsilon_1}+\bar{\alpha} \delta_{\epsilon_2}$$
Then using the $\cup$-convexity of $\epsilon \mapsto (1-2\epsilon)^{2n}$ we have
\begin{align*}
\underbrace{1-2\epsilon}_{\gamma_{\text{BEC},n}} \geq \underbrace{\alpha (1-2\epsilon_1)^{2n} + \bar{\alpha} (1-2\epsilon_2)^{2n}}_{\gamma_{a,n}} \geq \underbrace{(1-2\epsilon)^{2n}}_{\gamma_{\text{BSC},n}}
\end{align*}
\end{proof}

The polynomial $\rho$ is supposed to be increasing over $[0;1-2\epsilon]$, that is over a range that covers all the values the moments can take. Using this and \ref{l2}, the optimizers to each term in the series expansion of 
$
\Phi(\rho(a)) \stackrel{(\ref{phirhogen})}{=} \rho(1) - \sum_n a_{\Phi,n}\rho(\gamma_{a,n})
$
are the same, so we know they are the global optimizers.

\subsection{Proof of \ref{l1}}
\begin{proof}
We simply use $\gamma_{a,n} \leq \gamma_{a,1}$ and the monotonicity of $\rho$ to get
$$\forall n \, \rho(\gamma_{a,n}) \leq \rho(\gamma_{a,1})$$
Then 
\begin{align*}
\Phi(\rho(a)) \stackrel{(\ref{phirhogen})}{=} \rho(1) - \sum_n a_{\Phi,n}\rho(\gamma_{a,n}) \geq \rho(1) - \rho(\gamma_{a,1})\underbrace{\sum_n a_{\Phi,n}}_{1},
\end{align*}
and using Lemma \ref{l1} and again the monotonicity of $\rho$ 
\begin{align*}
\rho(\gamma_{a,1}) \leq \rho(\gamma_{\text{BSC},1}) = \rho\big(f_{\Phi}^{-1}(\phi_0)^2\big)
\end{align*}
(\ref{b1}) follows.
\end{proof}

\section{Other Inequalities}\label{obounds}
Here we give other inequalities that can be derived using the power series expansion, just as in the proofs of (\ref{t1}) and (\ref{t2}). We will only prove (\ref{i5}) along with the equality case which is the second part of (\ref{prev}). Remember that $\Phi$ stands for either $H$ or $B$. The reals $\alpha$ and  $\beta$ sum to $1$.
%
{\small
\begin{align}
1-\Phi(a\boxtimes b) &\geq (1-\Phi(a))(1-\Phi(b)) \label{i5}\\
\Phi(a^{\boxtimes d}) \leq \Phi(a)\Big(d -& \Phi(a) - \Phi(a^{\boxtimes 2}) - \ldots - \Phi(a^{\boxtimes d-1}) \Big) \label{i*}\\
1-\Phi(a) &\leq \sqrt{1-\Phi(a\boxtimes a)}\label{i2}\\
1-\Phi(a \boxtimes b) &\leq \sqrt{1-\Phi(a\boxtimes a)}\sqrt{1-\Phi(b \boxtimes b)}\label{i3} \\
1-\Phi(a \boxtimes b) &\leq \sqrt{1-\Phi(a\boxtimes a\boxtimes b)}\sqrt{1-\Phi(b)}\label{i4} \\
\Phi \left((\alpha a + \beta b)^{\boxtimes d} \right)&\geq \alpha \Phi(a^{\boxtimes d}) + \beta \Phi(b^{\boxtimes d})\label{i18}\\
\sqrt[d]{1-\Phi((\alpha a + \beta b )^{\boxtimes d})} &\leq \alpha\sqrt[d]{1-\Phi(a^{\boxtimes d})} +\beta \sqrt[d]{1-\Phi(b^{\boxtimes d})} \label{i6}\\
\Phi(a) &\leq f_\Phi(1-2E(a))\label{i1}\\
1-\Phi(a \boxtimes b) &\leq (1-\Phi(a))(1-2E(b))\label{i7}
\end{align}
}
\begin{proof}
(\ref{i5}): We do the same as in \ref{pt1}, except using another inequality than Jensen. Recall from (\ref{e1}) that
$$1-\Phi(a \boxtimes b) = \sum a_{\Phi,n} \gamma_{a,n} \gamma_{b,n}.$$
We use the following corollary of FKG inequality
$$\field{E}(fg) \geq \field{E}(f)\field{E}(g),$$
whenever $f$,$g$ have the same monotonicity. Equality case is when $f$ or $g$ is constant a.e. . Here $f:n \mapsto \gamma_{a,n}$, $g:n\mapsto \gamma_{b,n} $ and $\field{E}(f)=\sum a_{\Phi,n} f_n$. 

So, since the moments are decreasing, we get 
\begin{align*}
1-\Phi(a \boxtimes b) &= \sum a_{\Phi,n} \gamma_{a,n} \gamma_{b,n} \\
&\geq \sum a_{\Phi,n} \gamma_{a,n} \sum a_{\Phi,n} \gamma_{b,n}\\
&= (1-\Phi(a))(1-\Phi(b))
\end{align*}
with equality when $a$ or $b$ is from the $BEC$ family.
%
\end{proof}

\section{An application : studying the area threshold}\label{sareat}
Remember our initial problem which was to study when (\ref{area}) holds.
Fix $c_0>0$, we would like to know first, when{\small
\begin{align}\label{targ2}
-A = - h - (d_l-1-\quo)H(a^{\boxtimes d_r})+(d_l-1)H(a^{\boxtimes d_r-1}) \geq c_0
\end{align}
}
holds. We are going to show
\begin{lem}\label{neglem}
If the following two conditions are fulfilled then (\ref{targ2}) holds.
\begin{itemize}
\item[(i)] $\big(1-2h_2^{-1}(h)\big)^2 \leq \big(\frac{c_0}{d_l-1}\big)^{\frac{1}{d_r-1}}$
\item[(ii)] $h\leq \frac{d_l}{d_r}-2c_0 $
\end{itemize}
\end{lem}
\begin{proof}
Define 
\begin{align*}
d & =d_r & \kappa &= \frac{d_l-1-\quo}{d_l-1} &\rho &= X^{d-1}-\kappa X^{d}
\end{align*}

We are going to use the bound from Lemma \ref{l1}. 
The condition for $\rho$ to be increasing over the range of interest is
$\big(1-2h_2^{-1}(h)\big)^2 \leq \frac{d_r-1}{\kappa d_r}$ 
which is always true when $\kappa$ is given the value $\kappa = \frac{d_l-1-\quo}{d_l-1}$. So by Lemma \ref{l1}
{\small
\begin{align}\label{b12}
H(\rho(a)) \geq 1-\kappa - \left(1-2h_2^{-1}(h)\right)^{2(d-1)} + \kappa \left(1-2h_2^{-1}(h)\right)^{2d}
\end{align}
}
and then,

{\small
\begin{align*}
&- h - (d_l-1-\quo)H(a^{\boxtimes d_r})+(d_l-1)H(a^{\boxtimes d_r-1}) \\ 
&= - h+(d_l-1)H(\rho(a)) \\
&\stackrel{(\ref{b12})}{\geq} -h+(d_l-1)\Big[1-\kappa-\rho((1-2h_2^{-1}(h))^2\big)\Big]\\
\end{align*}
}
Also $(d_l-1)(1-\kappa) = \quo$. So for (\ref{targ2}) to hold it is enough that 
\begin{align*}
&-h+(d_l-1)\Big[1-\kappa-\rho((1-2h_2^{-1}(h))^2\big)\Big] \geq c_0 \\
&\Leftrightarrow \frac{d_l}{d_r}-h - (d_l-1)\rho((1-2h_2^{-1}(h))^2\big) \geq c_0
\end{align*}
which can be rewritten
\begin{align}\label{int1}
\frac{d_l}{d_r}-h \geq \underbrace{(d_l-1)\rho((1-2h_2^{-1}(h))^2\big)}_{\xi(h)} +c_0
\end{align}
(i) is s.t. $\xi(h) \leq c_0$, and then (ii) makes (\ref{int1}) true. Indeed,
{\small
\begin{align*}
\xi(h) &= (d_l-1)(1-2h_2^{-1}(h))^{2d_r-2}- (d_l-1-\frac{d_l}{d_r})(1-2h_2^{-1}(h))^{2d_r} \\
&\leq (d_l-1)(1-2h_2^{-1}(h))^{2d_r-2}\\
&\leq (d_l-1)\Big(\frac{c_0}{d_l-1}\Big)^{\frac{d_r-1}{d_r-1}} = c_0
\end{align*}
}
If we are interested only in the sign of $A(h)$ and not how far it is from $0$, we can let $c_0 = f(d_l,d_r)$ to increase the range of valid $h$. For instance, taking $c_0 = (d_l-1) \exp(-\sqrt{d_r-1})$, 
\begin{align*}
(i) \Leftrightarrow h_2^{-1}(h) \geq \frac{K}{\sqrt{d_r}}(1 + o(1))
\end{align*}
Where $K$ is some constant. Asymptotically this can be taken (changing the constant) to be simply
$$
h_2^{-1}(h) \geq \frac{K}{\sqrt{d_r}}.
$$
In the end, we are left with
\begin{prop}\label{cclarea}
For, $d_l,d_r$ large enough, the range for which (\ref{area}) holds contains an interval of the form $[L(d_l,d_r);R(d_l,d_r)]$ where 
\begin{align*}
L(d_l,d_r)&= h_2(\frac{K}{\sqrt{d_r}}) \\
R(d_l,d_r)&= \frac{d_l}{d_r} - o(d_r \exp(-\sqrt{d_r}))
\end{align*}
\end{prop}
\begin{rem}
Actually, changing $c_0(d_l,d_r)$, we could replace any $\sqrt{.}$ by $(.)^{\alpha}$ for any $\alpha <1$.
\end{rem}
\end{proof}

Proposition \ref{t1} in this context can be rephrased as
\begin{cor}
Define $\kappa$ as above
$$
\kappa = \frac{d_l-1-\quo}{d_l-1}
$$
Amongst channels $a$, with fixed entropy h, assuming the following condition is fulfilled
\begin{align}\label{cond}
(1-2h_2^{-1}(h))^{2} \leq \frac{\kappa-2}{d_r\kappa}
\end{align}-
then the BEC(h) minimizes
$$
-h-(d_l-1-\frac{d_l}{d_r})H(x^{\boxtimes d_r}) + (d_l-1)H(x^{\boxtimes d_r-1})
$$
\end{cor}
\section{Convex optimization and the shape of extremal densities}\label{conv}
Classical convex analysis provides powerful tools that allow - at least in the case where the target functional is linear - to drastically reduce the range of possible optimizers. Remember we represented the channels by probability measures over $[0;1]$. The basic principle is as follows

\begin{thm}[Dual Caratheodory]\label{car}
Take $\Phi$ \emph{any} continuous linear functional over BMS channels, like all those discussed above, and consider the following problem
\begin{align*}
\mbox{OPT }&\Phi(a)\\
\mbox{s.t. }&(\Phi_1(a),\ldots, \Phi_m(a)) = (\phi_1,\ldots,\phi_m)\\
\end{align*}
Then there are extremal densities $a_{+}$ and $a_{-}$ with support of cardinality at most $m+1$.
\end{thm}

\begin{disc}
The constraints are also assumed to be linear. A more extensive source on the topic is \cite{conv}.

This principle sheds some light on the fact that the \text{BSC} (which has one mass point in our representation) and the \text{BEC} (which has two) appear so often as extremal densities, when we consider problems with a single constraint. Indeed, one constraint corresponds to at most two mass points.

Extensions of the Caratheodory Principle were amongst the tools used in \cite{EIC07} to track \emph{two} channel parameters (namely $H$ and $E$) through the process of iterative decoding. As a result new bounds on iterative decoding were shown.
\end{disc}
%

It seems hard to derive proofs using solely \ref{car}. However it can be used to do numerical experiments. One way to proceed is as follows. Consider the target functional $\Phi(\rho(a))$ where $\rho$ is of degree $d$. Introduce $d$ variables $a_1,a_2,\ldots,a_d$ and replace (for $k\leq d$)
\begin{align*}
\Phi(a^{\boxtimes k}) \rightsquigarrow {d \choose k}^{-1}\sum_{1\leq i_1<\ldots < i_k\leq d} \Phi(a_{i_1}\boxtimes \ldots\boxtimes a_{i_k})
\end{align*} 

Denote $\tilde{\Phi}(a_1,\ldots,a_d)$ the expression we get. If it is maximized by a tuple where all coefficients $a_i$ are the same, then we know the initial expression has the same maximizer. To maximize $\tilde{\Phi}$, a simple tractable heuristic is optimizing coordinate after coordinate. Starting from random $a_i$s, to fix each coordinate except coordinate $i$, then find the best combination of two $\delta's$ for this coordinate. And repeat for all $i\leq d$. This gave good results for the motivational expression of (\ref{area}) and led to the claim 

\begin{claim}
The expression in (\ref{area}), for any $h$ and when $(d_l,d_r) = (3,6)$ or $(5,10)$ (the cases we tested) is always minimized by the BSC and maximized by the BEC.
\end{claim}

\section*{Acknowledgment}
The author would like to thank R\"udiger Urbanke his guidance and support.

\bibliographystyle{IEEEtran}
\bibliography{refs}

\begin{thebibliography}{1}
\providecommand{\url}[1]{#1}
\csname url@samestyle\endcsname
\providecommand{\newblock}{\relax}
\providecommand{\bibinfo}[2]{#2}
\providecommand{\BIBentrySTDinterwordspacing}{\spaceskip=0pt\relax}
\providecommand{\BIBentryALTinterwordstretchfactor}{4}
\providecommand{\BIBentryALTinterwordspacing}{\spaceskip=\fontdimen2\font plus
\BIBentryALTinterwordstretchfactor\fontdimen3\font minus
  \fontdimen4\font\relax}
\providecommand{\BIBforeignlanguage}[2]{{%
\expandafter\ifx\csname l@#1\endcsname\relax
\typeout{** WARNING: IEEEtran.bst: No hyphenation pattern has been}%
\typeout{** loaded for the language `#1'. Using the pattern for}%
\typeout{** the default language instead.}%
\else
\language=\csname l@#1\endcsname
\fi
#2}}
\providecommand{\BIBdecl}{\relax}
\BIBdecl

\bibitem{mct}
T.~Richardson and R.~Urbanke, \emph{Modern Coding Theory}.\hskip 1em plus 0.5em
  minus 0.4em\relax Cambridge University Press, 2007.

\bibitem{EIC05}
S.~S. I.~Sutskover and J.~Ziv, ``Extremes of information combining,''
  \emph{IEEE Trans. Inform. Theory}, vol.~51, no.~4, 2005.

\bibitem{EIC03}
S.~H. I.~Land, P.~Hoeher and J.~B. Huber, ``Bounds on information combining,''
  \emph{Proc. of the Int. Conf. on Turbo Codes and Related Topics}, 2003.

\bibitem{URK11.2}
T.~R. S.~Kudekar and R.~Urbanke, ``Spatially coupled ensembles with belief
  propagation achieve capacity universally,'' \emph{Work in progress}, 2011.

\bibitem{conv}
A.~Barvinok, \emph{A Course in Convexity}.\hskip 1em plus 0.5em minus
  0.4em\relax AMS, 2002.

\bibitem{EIC07}
S.~S. I.~Sutskover and J.~Ziv, ``Constrained information combining: Theory and
  applications for ldpc coded systems,'' \emph{IEEE Trans. Inform. Theory},
  vol.~53, 2007.

\end{thebibliography}
\end{document}